\documentclass[aps,twocolumn,prb]{revtex4}
\usepackage{amsmath}
\usepackage{epsfig}

\begin{document}

\title{ Quantum pump driven fermionic Mach-Zehnder interferometer}
\author{S.-W. V. Chung$^{1,2}$, M. Moskalets$^{3}$, P. Samuelsson$^{4}$}
\affiliation{$^{1}$Department of Electronics Engineering and
Institute of Electronics, National Chiao-Tung University, Hsinchu 30010,
Taiwan\\
$^{2}$Department of Electrophysics, National Chiao-Tung University, Hsinchu
30010, Taiwan\\
$^{3}$Department of Metal and Semiconductor Physics, National Technical
University ''Kharkiv Polytechnic Institute'', 61002 Kharkiv, Ukraine \\
$^{4}$Division of Mathematical Physics, Lund University, S\"{o}lvegatan 14 A,
S-223 62 Lund, Sweden}
\date{\today }

\begin{abstract}
We have investigated the characteristics of the currents in a
pump-driven fermionic Mach-Zehnder interferometer. The system is
implemented in a conductor in the quantum Hall regime, with the two
interferometer arms enclosing an Aharonov-Bohm flux $\Phi$. Two
quantum point contacts with transparency modulated periodically in
time drive the current and act as beam-splitters. The current has a
flux dependent part $I^{(\Phi)}$ as well as a flux independent part
$I^{(0)}$. Both current parts show oscillations as a function of
frequency on the two scales determined by the lengths of the
interferometer arms. In the non-adiabatic, high frequency regime
$I^{(\Phi)}$ oscillates with a constant amplitude while the amplitude
of the oscillations of $I^{(0)}$ increases linearly with
frequency. The flux independent part $I^{(0)}$ is insensitive to
temperature while the flux dependent part $I^{(\Phi)}$ is
exponentially suppressed with increasing temperature. We also find
that for low amplitude, adiabatic pumping rectification effects are
absent for semitransparent beam-splitters. Inelastic dephasing is
introduced by coupling one of the interferometer arms to a voltage
probe. For a long charge relaxation time of the voltage probe, giving
a constant probe potential, $I^{(\Phi)}$ and the part of $I^{(0)}$
flowing in the arm connected to the probe are suppressed with
increased coupling to the probe. For a short relaxation time, with the
potential of the probe adjusting instantaneously to give zero time
dependent current at the probe, only $I^{(\Phi)}$ is suppressed by the
coupling to the probe.
\end{abstract}

\maketitle

\section{\protect\bigskip Introduction}

Phase coherence in solid state conductors is a property of fundamental
interest. The prospect of solid state quantum information has also put
the focus on possible applications based on phase coherence. With the
progress of mesoscopic physics it has become possible to
experimentally explore the properties of quantum phase coherence in
solid state conductors in a controllable way.\cite{Been91} As a
prominent example, the fermionic (electronic) analog of the well known
bosonic (optical) Mach-Zehnder interferometer(MZI) was recently
realized by Ji {\it et al} \cite{Ji03} and further investigated in
Refs. [\onlinecite{Neder06,Litvin}]. The absence of closed electron
orbits makes the MZI the most elementary interferometer and therefore
of particular interest.

The MZI experiments\cite{Ji03,Neder06,Litvin} were all implemented in
a conductor in the integer quantum Hall regime, where the electrons
propagate along unidirectional, quantum mechanical edge states and
quantum point contacts (QPCs) act as beam splitters. In the
experiments\cite{Ji03,Neder06,Litvin} the visibility of the
conductance oscillations as a function of flux $\Phi$ were reduced
below the ideal value, a signature of dephasing of the electrons
propagating along the edges. Dephasing in the MZI was investigated in
several theoretical works. Originally, Seelig and
B\"uttiker\cite{Seelig01} investigated the effect of dephasing on the
conductance oscillations due to Nyquist noise. Following the
experiment in Ref. [\onlinecite{Ji03}], where also the shot noise was
measured, a number of works investigated the effect of dephasing on
the current and the noise. The dephasing was introduced via
fluctuating classical
potentials\cite{Marquardt04a,Marquardt04b,Forster05} and by coupling
the MZI to a quantum bath\cite{Marquardt05} as well as to a voltage
probe.\cite{Marquardt04a,Marquardt04b,Vane05} Recently these studies
were extended to the full distribution of the transferred charge, both
for a fluctuating classical potential\cite{Forster05} as well as a
voltage probe\cite{Pilgram05,Forster06} as a source of dephasing.

Taken together, these theoretical investigations have provided a
qualitative picture of the effect of dephasing on transport properties
in the MZI. The experimental situation is however not conclusive. In
the very recent work by Litvin {\it et al} \cite{Litvin} the observed
temperature and voltage dependence of the visibility of the
conductance oscillations are in good agreement with the noninteracting
theory of Ref. [\onlinecite{Vane05}]. In particular, the effect of the
interferometer arm asymmetry is clearly manifested. The overall
visibility is however low, a couple of percent.  In contrast, in the
recent work by Neder {\it et al}\cite{Neder06} the visibility is high,
$\sim 60\%$, but the voltage dependence of the conductance visibility
was found to be insensitive to arm asymmetry, however showing a clear
lobe structure. A possible explanation for the findings of
Ref. [\onlinecite{Neder06}] was also suggested, invoking interactions
between electrons at different edge states.\cite{Sukh06} The
experimental situation thus motivates further investigations of the
coherent transport properties of the MZI.

In this work we propose to investigate the properties of the currents
in a pump driven MZI. In contrast to previous work, both experimental
and theoretical, all electronic reservoirs are kept at the same
potential. The current is instead created via the quantum pump effect,
\cite{Thouless,BPT,Spivak,Brouwer98,Zhou} by varying periodically the
transparencies of the two QPCs. Working in the adiabatic, low pump
frequency limit the system is kept close to equilibrium. This
minimizes the effect of inelastic dephasing and hence allows for a
more detailed investigation of the coherence properties.

Theoretically, a large number of investigations of various aspects of
quantum pumping have been carried out, a representative collection can
be found in
Refs. [\onlinecite{Thouless,BPT,Spivak,Brouwer98,Zhou,Shutenko,AEGS00,VAA01,WWG02,MB02,PB03,AEGS04,
MMLM04,ZLCMcK04,SH04,CTCC04,GTFH05,CKS05,Benjamin06}]. However, only a
few experiments aimed at investigating quantum pumping of electrical
currents have been performed.\cite{SMCG99,DiCarlo03} In the MZI the
current is a true quantum interference effect. In addition, the
elementary structure of the MZI and the fact that the potential
applied at the QPC control both the pump effect and the scattering
properties of the QPCs makes the MZI a promising candidate for a
quantum pump. Previous studies of pumping in mesoscopic
interferometers have concerned Aharonov-Bohm,\cite{SH04,Citro06,Kim06}
double slit-quantum dot\cite{ZLCMcK04} and
two-particle\cite{SamButt05} interferometers, however, to the best of
our knowledge, not MZIs.

We use a Floquet scattering approach to the quantum pump problem.
\cite{MosButt02,Kim,MosButt04} This allows us to calculate the
currents in the MZI for arbitrary pumping strength, frequency and
temperature. In the Floquet picture, currents arise due to photon
assisted interference. It is found that the pumped current contains
both an Aharonov Bohm flux dependent part $I^{(\Phi)}$, due to
interfering paths enclosing the flux, and a flux independent part
$I^{(0)}$. Both current parts depend linearly on the pump frequency in
the low frequency, adiabatic regime and show oscillations as a
function of frequency in the high frequency, non-adiabatic regime. The
oscillations in the non-adiabatic regime occur on two different
frequency scales, governed by the interferometer arm length difference
and the mean arm length respectively. For the flux dependent current
$I^{(\Phi)}$, the oscillations have a constant amplitude while the
amplitude of the $I^{(0)}$ oscillations increases linearly with
frequency. The two current parts also display a different dependence
on temperature; the flux independent part is insensitive to
temperature while the flux dependent part is monotonically suppressed
with increasing temperature.

In the experiments in Refs. [\onlinecite{SMCG99,DiCarlo03}]
rectification effects made it difficult to distinguish the pumped
current. Importantly, in the MZI it is found that in the regime of low
amplitude, adiabatic pumping, rectification effects\cite{Brouwer01}
are absent for semitransparent beam-splitters. In order to investigate
the effect of dephasing on the pumped current, we consider one of the
interferometer arms connected to a voltage probe. Electrons injected
into the probe scatter inelastically and hence lose phase coherence
before being emitted out of the probe again. Two limiting regimes of
the charge relaxation, or RC, time of the voltage probe compared to
the pump period are considered; the long relaxation time regime, where
the potential of the probe is constant during the measurement, and the
short relaxation time regime where the potential of the probe adjusts
instantaneously in order to keep zero time dependent current at the
probe. In the long time regime the flux dependent current $I^{(\Phi)}$
as well as the part of the flux independent $I^{(0)}$ flowing in the
arm connected to the probe are successively suppressed by increasing
the dephasing, i.e. the strength of the coupling to the probe. In the
short time regime, only $I^{(\Phi)}$ is suppressed by dephasing.
 
The rest of the paper is organized as follows. In Sec. II the Floquet
scattering approach is first presented for an arbitrary mesoscopic
scatterer and then applied to the MZI. In Sec. III the properties of
the pumped currents are analyzed. Next, in Sec. IV the effects of
dephasing are investigated for different response times of the probe.
Finally, in Sec. V we conclude.

\section{Theory and Model}

\subsection{Floquet scattering approach}

For completeness we first briefly review the Floquet scattering
approach to pumping in mesoscopic conductors.
\cite{MosButt02,MosButt04} A mesoscopic system connected to $N$
reservoirs via single channel leads is considered. The system is
perturbed by some time-dependent parameters which all vary with the
same frequency $\omega$. The current flowing in the system in response
to the time-periodic perturbation is periodic in time.  Expanding the
current $I_{\alpha}(t)$ at reservoir $\alpha $ into a Fourier series,
we have
\begin{eqnarray}
I_{\alpha }\left( t\right) &=&\sum_{l=-\infty }^{\infty }\exp \left(
-il\omega t\right) I_{\alpha ,l},  \notag \\
I_{\alpha ,l} &=&\int_{0}^{\mathcal{T}}\frac{dt}{\mathcal{T}}\exp \left(
il\omega t\right) I_{\alpha }\left( t\right) ,
\label{floqcomp}
\end{eqnarray}
where $\mathcal{T}=2\pi /\omega $ is the period of the oscillations.
The Fourier component $I_{\alpha ,l}$ can be written \cite{Butt92}
\begin{equation}
I_{\alpha ,l}=\frac{e}{h}\int_{0}^{\infty }dE\left[ \left\langle \hat{b}%
_{\alpha }^{\dagger }\left( E\right) \hat{b}_{\alpha }\left( E_{l}\right)
\right\rangle -\left\langle \hat{a}_{\alpha }^{\dagger }\left( E\right) 
\hat{a}_{\alpha }\left( E_{l}\right) \right\rangle \right],  \label{I_l}
\end{equation}
with $\langle ... \rangle$ denoting a quantum statistical
average. Here $E_{l}=E+l\hbar \omega $ and $\hat{b}_{\alpha }\left(
E\right) $ and $\hat{a}_{\alpha }\left( E\right) $ are annihilation
operators for particles coming into and going out from the reservoirs
respectively. The operators $\hat{b}_{\alpha }\left( E\right) $ and
$\hat{a}_{\alpha }\left( E\right) $ are related via the Floquet
scattering matrix $s_{F}$ as
\begin{equation}
\hat{b}_{\alpha }\left( E\right) =\sum_{\beta =1}^{N}\sum_{n=-\infty
}^{\infty }s_{F,\alpha \beta }\left( E,E_{n}\right) \hat{a}_{\beta }\left(
E_{n}\right) ,  
\label{out_ope}
\end{equation}
where the element $s_{F,\alpha \beta }\left( E,E_{n}\right)$ is the
amplitude for scattering of an electron from reservoir $\beta $ at
energy $E_n$ to reservoir $\alpha $ and energy $E$. All the reservoirs
are in thermal equilibrium, giving the average
\begin{equation}
\left\langle \hat{a}_{\alpha }^{\dagger }\left( E_{n}\right) \hat{a}_{\beta
}\left( E_{m}\right) \right\rangle =f_{\alpha }\left( E_{n}\right) \delta
_{\alpha \beta }\delta _{nm},
\label{qstat}
\end{equation}
where $f_{\alpha }\left( E_{n}\right) =\left\{ 1+\exp
\left[E_{n}/k_{B}T_{\alpha }\right] \right\} ^{-1}$ is the Fermi
distribution function with $T_{\alpha }$ the temperature of reservoir
$\alpha$ and $k_{B}$ is the Boltzmann constant.

Substituting Eq. (\ref{out_ope}) into Eq. (\ref{I_l}) and taking into
account the unitarity of the Floquet scattering matrix,
\cite{MosButt04}
\begin{equation}
\sum_{\beta =1}^{N}\sum_{n=-\infty }^{\infty }s_{F,\alpha \beta
}^{\ast }\left( E,E_{n}\right) s_{F,\gamma \beta }\left(
E_{l},E_{n}\right) =\delta _{l,0}\delta_{\alpha,\gamma},
\label{unitarity}
\end{equation}
we can, with Eq. (\ref{qstat}), rewrite Eq. (\ref{I_l}) as
\begin{eqnarray}
I_{\alpha ,l} &=&\frac{e}{h}\int_{-\infty }^{\infty }dE\sum_{\beta
=1}^{N}\sum_{n=-\infty }^{\infty }\left[ f_{\beta }\left( E\right)
-f_{\alpha }\left( E_{n}\right) \right]  \notag \\
&&\times s_{F,\alpha \beta }^{\ast }\left( E_{n},E\right) s_{F,\alpha \beta
}\left( E_{n+l},E\right) .  \label{I_l_2}
\end{eqnarray}
Note that to get the above equation we have, compared to
Eq. (\ref{out_ope}), made the shift $E_{n}\rightarrow E$. At $l=0$ the
equation (\ref{I_l_2}) defines a dc current
\begin{eqnarray}
I_{\alpha ,dc} &=&\frac{e}{h}\int_{-\infty }^{\infty }dE\sum_{\beta
=1}^{N}\sum_{n=-\infty }^{\infty }\left| s_{F,\alpha \beta }\left(
E_{n},E\right) \right| ^{2}  \notag \\
&&\times \left[ f_{\beta }\left( E\right) -f_{\alpha }\left( E_{n}\right) %
\right] .  
\label{Idc}
\end{eqnarray}
\begin{figure}[t]
\includegraphics[width=0.4 \textwidth,angle=0]{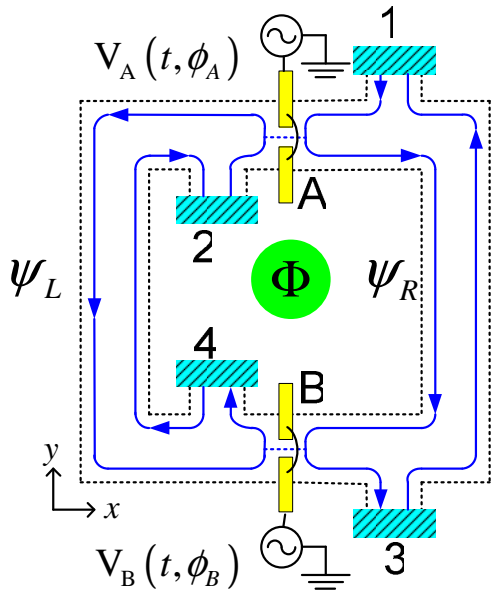}
\caption{The pump driven MZI implemented in a conductor in the quantum
Hall regime, supporting a single, unidirectional edge state. The
conductor is connected to four reservoirs $\alpha=1$ to $4$ kept at
the same potential. Two electrostatic split gates, at $A$ and $B$, are
acting as QPCs. The corresponding gate potentials $V_A(t,\phi_A)$ and
$V_B(t,\phi_B)$, with $\phi_A$ and $\phi_B$ the pumping phases, vary
periodically in time. The time dependent potentials give rise to
scattering in both real and energy space and are driving the pump
current. An Aharanov-Bohm flux $\Phi$ threads the MZI.}
\label{fig1}
\end{figure}
Through the rest of the paper we will focus on the dc-current.

\subsection{Mach Zehnder interferometer}

We consider a pump-driven Mach-Zehnder interferometer (MZI)
implemented in a conductor in the quantum Hall regime, as shown in
Fig. \ref{fig1}. Transport takes place along a single edge state
(filling factor one) and is unidirectional. Two electrostatic split
gates $A$ and $B$, defining quantum point contacts (QPCs) $j=A,B$, are
subjected to time dependent potentials
$V_j(t,\phi_j)=V_{sj}+V_j\cos(\omega t+\phi_j)$ with $\phi_j$ the
pumping phase. The pumping potentials give rise to scattering of
electrons between the edges as well as absorption or emission of one
or several quanta of energy $\hbar \omega$. An Aharonov-Bohm flux
$\Phi$ penetrates the interior of the interferometer. The conductor is
connected to four electronic reservoirs $\alpha=1$ to $4$. All four
reservoirs are kept at the same potential (grounded) and temperature
$T_{\alpha}=T$. Thus, the Fermi distribution functions for all the
reservoirs are the same,
$f_{\alpha}\left(E\right)=f_{0}\left(E\right)$, and in the absence of
the pumping potentials there is no current flow.

The scattering at the QPCs A and B, taking place both in real space
and in energy space, can be described by the Floquet scattering
matrices
\begin{equation}
S_j(E_n,E_m)=\left(
\begin{array}{cc}
r_j\left( E_n,E_m\right) & t_j'\left( E_n,E_m\right) \\ t_j\left(
E_n,E_m\right) & r_j'\left( E_n,E_m\right)
\end{array}
\right),  
\label{Svj}
\end{equation}
with primed amplitudes for particles incident on the QPCs from left in
Fig. \ref{fig1}. The QPCs thus act as inelastic beam splitters. We
consider the scattering amplitudes to be independent on energy on the
scale of the pump frequency. Consequently, $S_j(E_n,E_m)\equiv
S_{j,n-m}(E)$ can be expressed in terms of the Fourier coefficients
for the corresponding frozen scattering matrix \cite{4} $S_{j}\left(
E,t\right)$ as
\begin{eqnarray}
S_{j,n-m}\left(E \right)= \int_{0}^{\mathcal{T}}\frac{dt}{\mathcal{T}}e^{i\left( n-m\right)
\omega t} S_{j}\left( E,t\right) .
\label{Svj_nm}
\end{eqnarray}
Moreover, it is assumed that the scale of the energy dependence of the
QPC scattering amplitudes is much larger than the thermal energy
$k_BT$, allowing us to neglect the energy dependence of the Floquet
scattering matrix of the QPCs all together, $S_{j,n-m}\left(
E\right)=S_{j,n-m}$.

Propagating ballistically along the edges between the QPCs the
electrons pick up a phase containing both a geometrical part $k_mL_i$
and a part $\psi_i$ due to the Aharonov-Bohm flux, with $i=L,R$. Here
$\psi_{L}+\psi_{R}=2\pi \Phi/\Phi_{0}$ where $\Phi_{0}=h/e$ is the
flux quantum. It is assumed that the wavenumber $k_m=k(E_m)$ can be
taken linear in energy, \cite{Fertig88,Vane05}
\begin{equation}
k_{m}L_{i}=\zeta _{i}\left( \mu \right) +\frac{L_{i}}{\hbar v_{D}}%
\left( E+m\hbar \omega \right) ,
\end{equation}
where $\zeta _{i}\left( \mu \right)$ is the accumulated phase at the
Fermi energy and $v_{D}$ the drift velocity of the edge states. The
lengths of the interferometer arms are $L_L$ and $L_R$ respectively,
where we without loss of generality take $L_L \geq L_R$. The total
Floquet scattering amplitude can thus be expressed in terms of the
scattering amplitudes of the QPCs and the phases acquired along the
interferometer arms. For scattering from energy $E$ at reservoir $1$
to energy $E_n$ at reservoir $3$ the amplitude is
\begin{eqnarray}
s_{F,31}\left(E_{n},E\right) &=&\sum_{m=-\infty }^{\infty }\left[
r_{B,n-m}e^{ik_{m}L_{R}-i\psi _{R}}r_{A,m} \right.  \notag \\
&& \left. + t_{B,n-m}'e^{ik_{m}L_{L}+i\psi
_{L}}t_{A,m} \right] 
\label{s_41}
\end{eqnarray}
and similarly for the other amplitudes. Inserting these scattering
amplitudes into Eq. (\ref{Idc}) we arrive at the expression for the dc
current.

To perform an analysis of the entire parameter space, in the plots we
model for simplicity the QPC potentials with oscillating delta
function potentials
\begin{equation}
V_{j}\left(t,\phi_{j}\right)=\delta\left(x\right)\left(V_{sj}+2V_{j}\cos
(\omega t\ +\phi _{j})\right).
\end{equation}
We note that this choice leads to completely symmetric scattering
matrices, $t_{j,n}=t_{j,n}'$ and $r_{j,n}=r_{j,n}'$. It is pointed out
explicitly in the text below where this additional symmetry
qualitatively affects the result. The frozen scattering amplitudes of
the QPCs are given by
\begin{eqnarray}
t_j(t,\phi_j)=\frac{1}{1+im_e/(\hbar^2k_{\mu})[V_{sj}+2V_j\cos(\omega t+\phi_j)]},
\end{eqnarray}
and $r_j(t,\phi_j)=t_j(t,\phi_j)-1$, with $m_e$ the effective electron
mass and $k_{\mu}$ the wavenumber at the Fermi energy. This gives from
Eq. (\ref{Svj_nm}) the Fourier coefficients
\begin{eqnarray}
t_{j,n}&=&\frac{e^{-in \phi _{j}}}{\sqrt{\left[
1+ia_{j}\right]^{2}+b_{j}^{2}}} \notag \\ & \times& \left\{
\frac{i}{b_j}\left[1+ia_j-\sqrt{(1+ia_j)^2+b_j^2}\right]
\right\}^{\left| n\right|}, \notag \\ r_{j,n}&=&t_{j,n}-\delta_{n,0},
\label{tnm_rnm}
\end{eqnarray}
with $a_{j}=V_{sj}m_e/(\hbar^2 k_{\mu})$ and
$b_{j}=2V_{j}m_e/(\hbar^2 k_{\mu})$.

\section{Pumped current}

\begin{figure}[t]
\includegraphics[width=0.47\textwidth,angle=0]{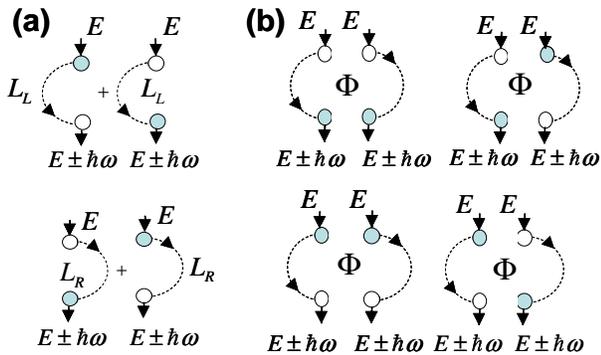}
\caption{Two qualitatively different types of first order
photon-assisted interference processes contributing to the current:
(a) along the same spatial paths L or R and (b) along the different
spatial paths L and R. The paths in (b) are sensitive to the enclosed
flux $\Phi$. Filled balls indicate inelastic scattering, the electrons
pick up or lose one quantum of energy $\hbar \omega$, while empty
balls indicate elastic scattering.}
\label{fig2}
\end{figure}

In the Floquet scattering picture, the pumping current arises due to
interference between different paths of the electrons in energy space,
i.e. photon-assisted interference.\cite{BM06} Due to the absence of
closed orbits in the MZI, there are only two different types of
interfering paths; the two paths either go along the same
interferometer arm, L or R, or along different arms. The latter paths
enclose the flux $\Phi$ in and give rise to an Aharonov-Bohm effect in
the pumped current. In Fig. \ref{fig2} different interfering paths
contributing to the current are shown. We note that an Aharonov-Bohm
effect in the pumped current was also predicted for other
interferometers.\cite{SH04,ZLCMcK04,Citro06,Kim06}

It is thus natural to part the total current into a flux dependent and
a flux independent part. Focusing on the current at reservoir $3$, we
write
\begin{equation}
I_{3,dc}=I_{3}^{\left( 0\right) }+I_{3}^{\left( \Phi \right)}.
\end{equation}
Inserting the scattering amplitudes in Eq. (\ref{s_41}) into the
current expression Eq. (\ref{Idc}) and carrying out the energy
integrals we arrive at the flux independent part
\begin{eqnarray}
&&I_{3}^{\left( 0\right) }= \frac{e\omega }{2\pi }\sum_{n=-\infty
}^{\infty}\sum_{m=-\infty }^{\infty }\sum_{p=-\infty }^{\infty }n
\nonumber \\
&&\times\left\{\left(r_{A,m}r_{A,p}^*+t_{A,m}'t_{A,p}^{\prime \ast}\right)t_{B,n-m}'t_{B,n-p}^{\prime \ast}
\right. \nonumber \\ 
&&\times\left. \exp\left[i\omega(m-p)\left(\tau+\frac{\hbar}{2E_c}\right)\right] \right. \nonumber \\
&&+\left. \left(r_{A,m}'r_{A,p}^{\prime \ast}+t_{A,m}t_{A,p}^{*}\right)r_{B,n-m}r_{B,n-p}^*
\right. \nonumber \\ 
&&\times\left. \exp\left[i\omega(m-p)\left(\tau-\frac{\hbar}{2E_c}\right)\right] \right\}
\label{I30}
\end{eqnarray}
and the flux dependent part
\begin{eqnarray}
&&I_{3}^{\left( \Phi \right) } =\frac{2eE_{c}}{h}g\left( T\right)
\sum_{n=-\infty }^{\infty }\sum_{m=-\infty }^{\infty }\sum_{p=-\infty
}^{\infty }\sin \left( \frac{n\hbar \omega }{2E_{c}}\right) \nonumber \\ 
&&\times 2\Re \left\{r_{B,n-m}t_{B,n-p}^{\prime \ast}\left(r_{A,m}t_{A,p}^*+t_{A,m}'r_{A,p}^{\prime \ast}\right)
\right.  \nonumber \\ 
&&\times \left. \exp\left[i\left(\psi_{LR}+\omega\left[(m-p)\tau+(n-p-m)\frac{\hbar}{2E_c}\right]\right) \right]\right\}, \nonumber \\
\label{I3phi}
\end{eqnarray}
where 
\begin{equation}
g\left( T\right) =\frac{\pi k_{B}T}{E_{c}}\mbox{csch}\left( \frac{\pi
k_{B}T}{E_{c}}\right),
\label{gfcn}
\end{equation}
the phase $\psi _{LR}=\zeta _{L}\left( \mu \right) -\zeta _{R}\left(
\mu \right) -2\pi \Phi /\Phi _{0}$ and $\Re$ denoting the real
part. 

In order to explicitly display the relevant energy and time scales we
have introduced the asymmetry energy $E_{c}=\hbar v_{D}/\left(
L_{L}-L_{R}\right)$ and the average time $\tau=(L_L+L_R)/(2 v_D)$ for
ballistic propagation between the QPCs. By definition
$\tau>\hbar/(2E_c)$. There are thus three different, possible pumping
regimes depending on the relation between the pump frequency $\omega$
and the frequency scales $E_c/\hbar$ and $1/\tau$: (i) For $\omega \ll
1/\tau$ the pumping is adiabatic, the total scattering amplitudes of
the MZI are independent on energy on the scale of the pumping
frequency $\omega$. For non-adiabatic pumping there are in addition
two regimes. (ii) In the intermediate frequency regime $1/\tau \ll
\omega \ll E_c/\hbar$ the pumped current is independent on the
interferometer asymmetry but depends on the total time $\tau$. (iii)
For $E_c/\hbar \ll \omega$, in the high frequency regime, the pumped
current depends both on the asymmetry and the total time.

Several important observations can be made directly from the formal
expressions in Eqs. (\ref{I30}) and (\ref{I3phi}). First, the flux
independent current $I_3^{(0)}$ is an incoherent sum of the currents
pumped through the left and right arms. The two currents are denoted
$I_{3L}^{(0)}$ [upper term in Eq. (\ref{I30})] and $I_{3R}^{(0)}$ (lower term)
respectively. Each current term, $I_{3L}^{(0)}$ or $I_{3R}^{(0)}$,
depends explicitly on the time for ballistic propagation through the
corresponding left or right arm, $\tau+\hbar/(2E_c)=L_L/v_D$ and
$\tau-\hbar/(2E_c)=L_R/v_D$. For the flux dependent current, no such
partitioning into left and right arm currents is possible.

Second, while the flux independent current $I_3^{(0)}$ is independent
on the temperature, the flux dependent current $I_3^{(\Phi)}$ is
monotonically suppressed with increasing temperature. Despite the fact
that both terms are of interference nature, they thus depend on
temperature in very different ways. The energy scale of the decay of
$I_3^{(\Phi)}$ is set by the asymmetry energy $E_c$: the factor $g(T)$
is equal to unity for $k_BT \ll E_c$ and decays as $\exp(-\pi k_B
T/E_c)$ for $k_BT \gg E_c$. This is qualitatively similar to the
voltage biased MZI\cite{Vane05} and can be understood as an effect of
energy averaging. Notably, the temperature dependence of the current
is affected neither by the pumping frequency $\omega$ nor by the
average time $\tau$.

\begin{figure}[t]
\includegraphics[width=0.4 \textwidth,angle=0]{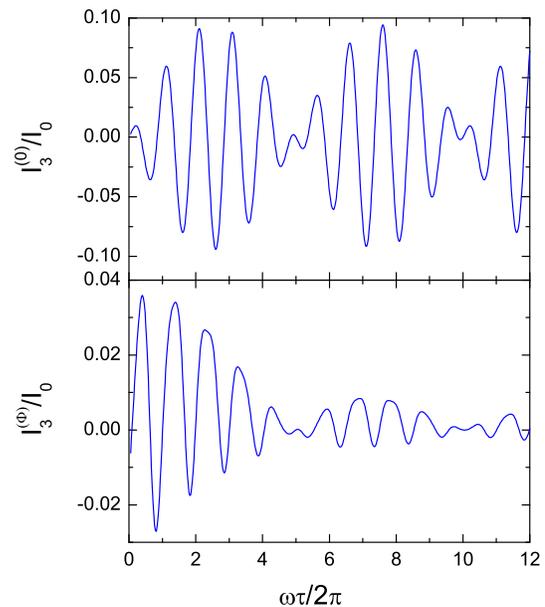}
\caption{The flux independent current $I_3^{(0)}$ (upper panel) and
the flux dependent current $I_3^{(\Phi)}$ (lower panel) as a function
of pump frequency. Guided by the experiments in
Ref. \onlinecite{Litvin,Neder06} we have taken an asymmetry,
$\tau=5\hbar/E_c$, and symmetric static beam splitters,
$a_A=a_B=1$. The other parameters are $b_A=0.4,b_B=1.3$ (strong
pumping), $\phi_A=0$, $\phi_B=0.8\pi$ and $\psi_{LR}=0.8\pi$.}
\label{fig3}
\end{figure}

Third, the qualitative behavior of the currents as a function of
frequency can also be understood from Eqs. (\ref{I30}) and
(\ref{I3phi}). It is clear that both currents $I_3^{(0)}$ and
$I_3^{(\Phi)}$ show oscillations in the non-adiabatic regime as a
function of $\omega$, on the scales $E_c/\hbar$ and $1/\tau$. In the
intermediate frequency regime (ii), for $1/\tau \ll \omega \ll
E_c/\hbar$, the pumped current is however insensitive to the asymmetry
and shows oscillations with the basic period $\tau$ only.  In regime
(iii), for $E_c/\hbar \ll \omega$, the pumped current shows
oscillations as a function of frequency on both the scales $1/\tau$
and $E_c/\hbar$. For a small asymmetry, $E_c \gg \hbar/\tau$, the
oscillations show a beating pattern with rapid oscillations on the
scale $1/\tau$ periodically modulated in amplitude on the scale
$E_c/\hbar$. This is illustrated in the plots in Fig. \ref{fig3}. We
point out that for the flux independent current $I_3^{(0)}$ the
beating pattern can simply be understood as the effect of adding the
two currents terms $I_{3L}^{(0)}$ and $I_{3R}^{(0)}$ with the two
different time periods $\tau\pm \hbar/(2E_c)$.

Importantly, the amplitudes of the oscillations of $I_3^{(0)}$ and
$I_3^{(\Phi)}$ show a different frequency dependence. As is clear from
Eq. (\ref{I30}), in the high frequency regime, $E_c/\hbar \ll \omega$,
the flux independent current is proportional to $\omega$ while the
flux dependent current has no frequency dependent prefactor. Plotting
the currents divided by the elementary pumped current
$I_0=e\omega/2\pi$, the oscillations of the normalized current
$I_3^{(0)}/I_0$ have a constant amplitude as a function of frequency
while the amplitude of $I_3^{(\Phi)}/I_0$ decreases as
$1/\omega$. This is illustrated in Fig. \ref{fig3}.

In the adiabatic regime, $\omega \ll 1/\tau$, both currents
$I_3^{(0)}$ and $I_3^{(\Phi)}$ in general show a linear dependence in
$\omega$. We however note that the our choice of a spatially symmetric
model potential for the QPCs leads to a flux independent part of the
current proportional to $\omega^2$. This quadratic frequency
dependence is not clearly visible in the plot in Fig. \ref{fig3}. This
sensitivity of $I_3^{(0)}$ to spatial symmetry is further discussed
below.

\subsection{Weak amplitude pumping}

Several of the properties of the pumped current become more
transparent in the limit of weak pumping, \cite{MosButt02} where only
one quantum of energy $\hbar \omega$ can be absorbed or emitted when
scattering through the MZI. This allows us to write the two current
parts on the form
\begin{eqnarray}
&&I_{3}^{\left( 0\right) }=\frac{-2e\omega }{\pi }\Im \left\{t_B'\delta
t_B^{*'}\left(\delta t_At_A^*+\delta
r_A'r_A^{\prime \ast}\right)\right. \nonumber \\
&\times& \left. \sin\left(\phi_A-\phi_B-\omega\left[\tau+\frac{\hbar}{2E_c}\right]\right) \right. \nonumber \\
&+& \left. r_B\delta r_B^*\left(\delta r_Ar_A^*+\delta
t_A't_A^{\prime \ast}\right) \right. \nonumber \\
&\times& \left. \sin\left(\phi_A-\phi_B-\omega\left[\tau-\frac{\hbar}{2E_c}\right]\right) \right\}
\label{I30weak}
\end{eqnarray}
with $\Im$ the imaginary part and
\begin{eqnarray}
&&I_{3}^{(\Phi)}=\frac{8eE_{c}}{h}g(T)\sin
\left(\frac{\hbar \omega }{2E_{c}}\right) \Im \left\{
e^{i\psi_{LR}} \right.\nonumber \\ 
&&\times \left. \left[\left(\delta r_Bt_B^{\prime \ast}+r_B\delta
t_B^{\prime \ast}\right)\left(r_A\delta t_A^*+\delta t_A'r_A^{\prime
\ast}\right) \right. \right. \nonumber \\ 
&&\left. \left. \times
\sin(\phi_A-\phi_B-\omega \tau) \right. \right. \nonumber \\
&&\left. \left.+r_Bt_B^{\prime \ast}\left[\delta r_A\delta t_A^*+\delta
t_A'\delta r_A^{\prime \ast}\right] \sin\left( \frac{\hbar \omega
}{2E_{c}}\right)\right] \right\}.
\label{I3phiweak}
\end{eqnarray}
Here we introduced the notation
$r_{j,0}=r_j,r_{j,0}'=r_j',t_{j,0}=t_j$ and $t_{j,0}'=t_j'$ for the
amplitudes to scatter elastically, without absorbing or emitting any
energy quantum, and $r_{j,\pm 1}=\delta r_j e^{\mp i\phi_j}, r_{j,\pm
1}'=\delta r_j' e^{\mp i\phi_j}, t_{j,\pm 1}=\delta t_j e^{\mp
i\phi_j}$ and $t_{j,\pm 1}'=\delta t_j' e^{\mp i\phi_j}$ for the
amplitudes to emit (-) or absorb (+) a single energy quantum.

Importantly, the terms in the current expressions directly correspond
to the first order scattering processes shown in Fig. \ref{fig2}. For
$I_3^{(0)}$, the upper term in Eq. (\ref{I30weak}), $I_{3L}^{(0)}$,
arises due to interference between electrons that propagate along the
left arm and pick up or lose a quantum $\hbar \omega$ at either $A$ or
$B$. These processes are shown at the top of panel (a) in
Fig. \ref{fig2}.  The lower term in Eq. (\ref{I30weak}),
$I_{3R}^{(0)}$, arises from the corresponding processes for electrons
propagating in the right interferometer arm.

For the flux dependent current $I_3^{(\Phi)}$, the interfering paths
go along different interferometer arms $L$ and $R$. The upper term in
Eq. (\ref{I3phiweak}) arises from processes where electrons pick up or
lose one quantum $\hbar \omega$ at different QPCs $A$ and $B$. These
processes are depicted to the right in panel (b). The lower term in
Eq. (\ref{I3phiweak}) arises from processes where both electrons pick
up or lose one quantum $\hbar \omega$ at the same QPC, $A$ or $B$,
depicted to the left in panel (b). Importantly, electrons which
scatter inelastically at the same QPC pick up the same information on
the pumping phase. Consequently, the corresponding interference term
is independent on the pumping phase, as seen in the lower term in
Eq. (\ref{I3phiweak}).

The weak amplitude expressions for the current also clearly
demonstrate the origin of the sign change of the current as a function
of frequency, as shown in Fig. \ref{fig3}. In the low frequency,
adiabatic limit the weak amplitude pumped current is always
\cite{Brouwer98} proportional to $\sin(\phi_A-\phi_B)$, i.e. the sign
of the current is determined by the pumping phase difference. In
Eqs. (\ref{I30weak}) and (\ref{I3phiweak}) the frequency formally
enters the current expressions as an additional pumping phase, thus
leading to an oscillating sign of the current as a function of
frequency.

\subsection{Adiabatic, weak pumping}

It is of particular importance to consider the weakly pumped currents
in the adiabatic limit, where the effects of inelastic dephasing are
minimized. In the adiabatic limit the current reduces to, using the
unitarity relations in Eq. (\ref{unitarity})
\begin{eqnarray}
I_{3,ad}^{\left( 0\right) }&=&\frac{ie\omega }{\pi}\sin(\phi_A-\phi_B) \left(r_B\delta r_B^*+r_B^*\delta r_B\right) \nonumber \\
&\times& \left(\delta t_At_A^*-\delta t_A't_A^{\prime \ast}-\delta
r_Ar_A^*+\delta
r_A'r_A^{\prime \ast}\right)
\label{I30weakad}
\end{eqnarray}
and
\begin{eqnarray}
I_{3,ad}^{(\Phi )}&=&\frac{2e\omega}{\pi}\sin(\phi_A-\phi_B) g(T) \Im \left\{e^{i\psi_{LR}} \right.\nonumber \\
&\times &\left. \left(\delta r_Bt_B^{\prime \ast}+r_B\delta t_B^{\prime \ast}\right)\left(r_A\delta t_A^*+\delta t_A'r_A^{\prime \ast}\right) \right\}.
\label{I3phiweakad}
\end{eqnarray}
Note that the second line in Eq. (\ref{I30weakad}) is purely
imaginary. From the expression of the flux independent current
$I_{3,ad}^{\left( 0\right)}$ we see explicitly the dependence on
spatial symmetry of QPC $A$. For a completely symmetric scattering
potential, i.e. primed scattering amplitudes equal to unprimed, the
adiabatic phase independent current is zero and the low frequency
current is $\propto \omega^2$. We point out that the absence of a
noticeable magnetic flux through the point contact area, i.e.
$t_A=t'_A$, is not enough to suppress the adiabatic current. We also
note that only the spatial symmetry of QPC $A$ is relevant, a
consequence of the chiral transport. That is,
reversing the sign of the quantum Hall magnetic field, the pumped
currents, now at reservoirs $1$ and $2$, would be sensitive to the
symmetry of QPC $B$ only.

From the dependence of $I_{3,ad}^{(\Phi)}$ on $\psi_{LR}$, as well as
the fact that $\psi_{LR}$ depends both on the Aharanov-Bohm flux as
well as phases picked up propagating along the edges [see definition
below Eq. (\ref{gfcn})], we can conclude that the flux dependent part
has no definite magnetic field symmetry. We also note that in the
adiabatic expression for the flux dependent current, the lower term in
Eq. (\ref{I3phiweak}), independent on the pumping phases, does not
contribute.

\subsection{Rectification effects}

For mesoscopic conductors, an unavoidable feature is stray
capacitances between the various circuit elements, i.e. the electronic
reservoirs, the electrostatic gates and the mesoscopic sample
itself. A capacitive coupling between the pumped QPC gates and the
electronic reservoirs induces an ac potential at the reservoirs (for
nonzero impedance of the current measurement circuit). This gives rise
to a rectification current which can obscure the pumped
current. \cite{SMCG99,Brouwer01,DiCarlo03,MMLM04,MosButt04} In
the MZI, for weak, adiabatic pumping, the rectified dc-current is in
the most general situation given by \cite{Brouwer01}
\begin{eqnarray}
I_{3,rect}&=&c_{A1}\frac{\partial G_{31}}{\partial V_A}+c_{B1}\frac{\partial G_{31}}{\partial V_B} \nonumber \\
&+&c_{A2}\frac{\partial G_{32}}{\partial V_A}+c_{B2}\frac{\partial G_{32}}{\partial V_B},
\end{eqnarray}
where the constants $c_{j\alpha}$ depend on the capacitive couplings ,
the impedance of the measurement circuit, the pumping phases and the
pumping amplitudes and $G_{\alpha\beta}=dI_{\alpha}/dV_{\beta}$ is the
conductance. From the theory for the conductance of the MZI in
Ref. [\onlinecite{Vane05}] we have
\begin{eqnarray}
\frac{\partial G_{31}}{\partial V_A}&=&\frac{\partial T_A}{\partial
V_A}\frac{\partial G_{31}}{\partial T_A}=\frac{\partial T_A}{\partial
V_A}\frac{2e^2}{h}\nonumber \\
&\times&\left(T_B-R_B+H\frac{R_BT_B(R_A-T_A)}{2\sqrt{R_AT_AR_BT_B}}\right)
\label{rect}
\end{eqnarray}
and similarly for the other conductance derivatives. Here
$T_A=1-R_A=|t_A|^2=|t_A'|^2$ is the transmission probability of the
static QPC $A$ and $H=H(k_BT,E_c,\Phi)$ a function dependent on the
different energy scales $k_BT$ and $E_c$ and the enclosed flux
$\Phi$. The rectification current and the pumped current thus depend
differently on the scattering parameters, the magnetic flux and the
energy scales, allowing one to distinguish experimentally between the
two currents. In particular, from Eq. (\ref{rect}) it is clear that
working with semitransparent beam splitters $T_A=R_A=1/2$ and
$T_B=R_B=1/2$ the rectification currents are zero. This holds
independently on the values of the individual couplings $c_{j\alpha}$.

We also emphasize that induced ac-potentials at the reservoirs do not
simply lead to a rectification current which is incoherently added to
the pumped current; there is in general also a current due to
interference between processes responsible for the pumped current and
the rectification current. \cite{MosButt04} However, the induced
ac-potential is proportional to\cite{Brouwer01} $dV_A/dt,dV_B/dt \sim
\omega$ and in the weak amplitude, adiabatic limit the interference
current is consequently $\propto \omega^2$.

A related issue is the effect of the temporary charging of the MZI
itself due to the pumping. In the calculations and discussions of the
pumped current above we have neglected this effect, i.e we have
considered noninteracting electrons. An interacting theory should also
take into account screening at the edges and the effect of capacitive
couplings of the edges to e.g. each other and to the electrostatic
gates. This would require a self-consistent determination of the time
dependent edge state potentials.\cite{Christen,PB98} Such an
interacting theory however goes beyond the scope of this paper.

\section{Dephasing}

\begin{figure}[h]
\includegraphics[width=0.4 \textwidth,angle=0]{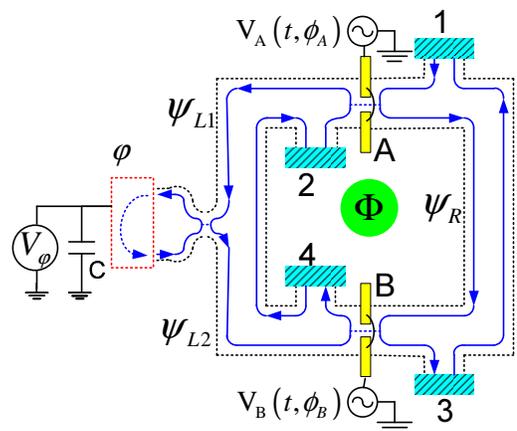}
\caption{The pump driven MZI of Fig. \ref{fig1} with the left arm
connected with strength $\epsilon$ to a voltage probe $\varphi$. The
dynamics of the potential $V_{\varphi}$ of the probe is governed by
the charge relaxation time (see text).}
\label{figdep}
\end{figure}

An important problem in the study of interference phenomena in
mesoscopic conductors is decoherence. Phase information of the
electrons propagating in the MZI is lost. Various approaches to
dephasing in the voltage biased MZI were discussed in the
introduction. Here we introduce dephasing in the MZI by coupling one
of the arms of the interferometer to a voltage probe as shown in
Fig. \ref{figdep}. A voltage probe is an additional electronic
reservoir with the potential left floating. Electrons entering the
voltage probe are incoherently fed back into the interferometer arm,
thereby suppressing phase coherence. Voltage probes as means to
introduce incoherent, inelastic scattering was proposed by
B\"uttiker.\cite{Butt86,Butt88} The concept has thereafter been
extended and applied to a large number of mesoscopic conductors, both
theoretically and experimentally. A recent account of this development
was given in Ref. [\onlinecite{BlanterButt00}].

For our purposes, in quantum Hall systems the theory of current and
noise in the presence of voltage probes was developed in
Ref. [\onlinecite{TexierButt92}] and applied to a voltage biased MZI
in Refs. [\onlinecite{Marquardt04a,Marquardt04b,Vane05}]. We point out
that the very recent experiments by Oberholzer {\it et
al},\cite{Ober06} investigating the current cross correlations in a
quantum Hall geometry coupled to a voltage probe, were in excellent
agreement with the theory of
Ref. [\onlinecite{TexierButt92}]. Moreover, dephasing of the pumped
current via voltage probes was considered in
Refs. [\onlinecite{MosButt01,Cremers02,PoliButt05}].

The elastic scattering matrix of the contact to the probe is given by
\begin{equation}
\left(
\begin{array}{cc} 
 \sqrt{1-\epsilon} & i\sqrt{\epsilon} \\ i\sqrt{\epsilon} & \sqrt{1-\epsilon}
\end{array}
\right).  
\label{Sdep}
\end{equation}
Here $\epsilon$ governs the strength of the dephasing. For
$\epsilon=1$ the dephasing is complete, i.e. the probe is fully
coupled to the MZI and all electrons propagating along the
interferometer arm enter the probe. For $\epsilon=0$ the transport is
fully coherent, the probe is decoupled from the MZI.

An important property of the voltage probe is the charge relaxation
time of the voltage probe,\cite{Pilgram05,Forster06} i.e. the time
scale on which the probe is charged or discharged. The charge
relaxation time determines the dynamics of the potential
$V_{\varphi}(t)$ of the probe and consequently the response to the
injected, time dependent charge. The charge relaxation time is given
by the RC-time $\tau_{RC}=RC$, with $R$ the charge relaxation
resistance and $C$ the capacitance (see Fig. \ref{figdep}). B\"uttiker
and one of the authors\cite{MosButt01} considered adiabatic pumping in
a conductor connected to a voltage probe, assuming instantaneous
charge conservation at the probe, i.e. a relaxation time much shorter
than the pump period. Cremers and Brouwer \cite{Cremers02}
investigated the pumped current in a chaotic quantum dot in the same
short relaxation time limit. Considering the experimental setup of
Ref. [\onlinecite{SMCG99}], Polianski and Brouwer\cite{PoliBrouw01}
investigated the adiabatic dynamics of the floating potential of
reservoirs. They considered the two limiting cases of long and short
relaxation time compared to the pump period. Here we will consider the
same limiting cases of short and long relaxation time for the voltage
probe, without the restriction to adiabatic pumping.

\subsection{Long charge relaxation time $\tau_{RC} \gg {\cal T}$}

First, the case with long relaxation time is considered, where the
potential of the probe does not react on the injected charge on the
time scale of the pumping period, $\tau_{RC} \gg \cal T$. In this
situation the potential of the probe is constant during the
measurement. Since the particles entering the probe have scattered at
the adiabatically pumped QPC A only, there is no dc-current flow into
the probe and the potential of the probe $V_{\varphi}$ stays at the
same potential as the four reservoirs of the MZI. We thus have an
extended pumping problem with five instead of four equipotential
reservoirs, which can be treated along the same lines as above.

First, the coupling of the MZI to the probe leads to a modification of
the scattering amplitudes in Eq. (\ref{s_41}), as $s_{F,\alpha\beta}
\rightarrow \tilde s_{F,\alpha\beta}$, with e.g.
\begin{eqnarray}
&& \tilde s_{F,31}\left(E_{n},E\right) =\sum_{m=-\infty }^{\infty }\left[
r_{B,n-m}e^{ik_{m}L_{R}-i\psi _{R}}r_{A,m} \right.  \notag \\
&& \left. + \sqrt{1-\epsilon}t_{B,n-m}'e^{ik_{m}L_{L}+i\psi
_{L}}t_{A,m} \right] 
\end{eqnarray}
and similar for the other amplitudes to scatter from reservoirs $1$
and $2$ to $3$ and $4$. Moreover, there are now the amplitudes to
scatter to and from the probe, as e.g. from $\varphi$ to $3$
\begin{eqnarray}
\tilde s_{F,3\varphi}\left(E_{n},E\right)
&=&i\sqrt{\epsilon}t_{B,n}'e^{ik_0L_{L2}},
\end{eqnarray}
where $L_{L2}$ is the length along the left edge between the probe and
QPC B (see Fig. \ref{figdep}). Inserting the scattering amplitudes
$\tilde s_{F,\alpha\beta}$ into the formula for the dc-current,
Eq. (\ref{Idc}), we arrive at the result that the coherent current is
modified as
\begin{eqnarray}
I_{3L}^{(0)} &\rightarrow & (1-\epsilon) I_{3L}^{(0)},\nonumber \\ 
I_{3}^{(\Phi)} &\rightarrow & \sqrt{1-\epsilon} I_{3}^{(\Phi)}.
\label{currlongdep}
\end{eqnarray}
The flux independent current in the arm to which the probe is coupled,
$I_{3L}^{(0)}$, is successively suppressed for increasing coupling
$\epsilon$ to the probe. For perfect coupling, $\epsilon=1$, the
current $I_{3L}^{(0)}$ is zero. The pumped current flowing through the
arm not connected to the probe is however unaffected by the coupling
to the probe. In contrast, the entire flux independent current
$I_{3}^{(\Phi)}$ is suppressed for increasing coupling to the probe,
down to zero for perfect coupling.

\subsection{Short charge relaxation time $\tau_{RC} \ll {\cal T}$}

In the limit of a response time much shorter than the pumping period,
$\tau_{RC} \ll {\cal T}$, the potential of the probe $V_{\varphi}(t)$
adjusts instantaneously, in order to keep the time dependent current
at the probe zero, $I_{\varphi}(t)=0$. This corresponds to that all
frequency components of the current [see Eq. (\ref{floqcomp})] are
zero,
\begin{eqnarray}
I_{\varphi,l} = 0.
\label{zerofloq}
\end{eqnarray}
Since the electrons entering the voltage probe are rapidly
thermalized, the electrons in the probe can be considered in the same
way as electrons in a reservoir with oscillating potential, i.e. in
dynamical equilibrium. The oscillating potential gives rise to a
nonequilibrium distribution of the electrons leaving the
probe. Formally, assuming a uniform potential of the probe, we follow
the scattering approach in Ref. [\onlinecite{PB98}] and introduce
annihilation operators for the electrons emitted from the probe as
\begin{equation}
\hat{a}_{\varphi}^{\prime}\left( E\right) =\sum_{n=-\infty }^{\infty
}L_{-n}\hat{a}_{\varphi}\left(E_n\right) .
\label{a_phi}
\end{equation}
Here the operators $\hat{a}_{\varphi }\left( E\right) $ describe
equilibrium electrons: $\left\langle \hat{a}_{\varphi }^{\dagger
}\left( E_{n}\right) \hat{a}_{\varphi }\left( E_{m}\right)
\right\rangle =f_{0}\left( E_{n}\right) \delta _{n,m}.$ The amplitudes
$L_{n} $ are defined as
\begin{equation}
L_{n}=\int_{0}^{\mathcal{T}}\frac{dt}{\mathcal{T}}\exp \left(in\omega
t\right) \exp \left( -i\int dt eV_{\varphi}(t)/\hbar
\right) .  
\label{Ln}
\end{equation}
The annihilation operators for particles injected into the probe can
then be written as
\begin{eqnarray}
&&\hat{b}_{\varphi}\left( E\right)=\sum_{n=-\infty }^{\infty }\left[s_{F,\varphi 1}(E,E_n)\hat a_1(E_n) \right. \nonumber \\
&& \left. +s_{F,\varphi 2}(E,E_n)\hat a_2(E_n)+\sqrt{1-\epsilon}L_{-n} \hat a_{\varphi}(E_n)\right].
\label{out_ope_phi}
\end{eqnarray}
Importantly, the amplitudes $L_n$ in Eq. (\ref{a_phi}) effectively
describe forward, inelastic scattering from the probe out into the
MZI. We can consequently combine the amplitudes $L_n$ for excitation
of the electrons in the probe and the amplitudes $\tilde
s_{F,\alpha\beta}$ for scattering in the MZI with zero probe potential
into a new, unitary Floquet scattering matrix $\bar
s_{F,\alpha\beta}$. This gives $\bar s_{F,\alpha\beta}(E,E_n)=\tilde
s_{F,\alpha\beta}(E,E_n)$ for $\beta \neq \varphi$ and $\bar
s_{F,\varphi\varphi}(E,E_n)=\sqrt{1-\epsilon}L_{-n}$ and similarly for
$\bar s_{F,3\varphi}$ and $\bar s_{F,4\varphi}$. It is then possible
to proceed as above and insert the scattering amplitudes $\bar
s_{F,\alpha\beta}$ into the formula for the fourier components of the
current, Eq. (\ref{floqcomp}). This gives
\begin{eqnarray}
I_{\varphi ,l} &=&\frac{e}{h}\int_{0}^{\infty }dE\sum_{m=-\infty }^{\infty }\left[f_{0}\left( E_{m}\right)-f_0(E)\right]  \nonumber \\
&&\times \sum_{\beta}\bar{s}_{F,\varphi \beta }^{\ast }\left( E,E_{m}\right) 
\bar{s}_{F,\varphi \beta }\left( E_{l},E_{m}\right), 
\end{eqnarray}
where $\beta$ runs over $1,2$ and $\varphi$. The requirement of
instantaneous current conservation, Eq. (\ref{zerofloq}), then
directly gives
\begin{eqnarray}
\sum_{\beta}\bar{s}_{F,\varphi \beta }^{\ast }\left( E,E_{m}\right) 
\bar{s}_{F,\varphi \beta }\left( E_{l},E_{m}\right)=0,
\end{eqnarray}
which in terms of the amplitudes $L_n$ can be written
\begin{equation}
L_{m}^{\ast }L_{m+l}=\frac{1}{\epsilon}\sum_{\beta
=1,2}\bar{s}_{F,\varphi \beta }^{\ast }\left( E,E_{-m}\right)
\bar{s}_{F,\varphi \beta }\left( E_{l},E_{-m}\right).
\label{Ln_S}
\end{equation}
We will then use Eq. (\ref{Ln_S}) to calculate the dc-current at
reservoir $3$. The dc-current is given by Eq. (\ref{Idc}), now with
the scattering amplitudes $\bar s_{F,3\beta}$. Via the amplitude $\bar
s_{F,3\varphi}$ the current depends on the product
$L_m^*L_{m+l}$. Inserting the expression for $L_m^*L_{m+l}$ from
Eq. (\ref{Ln_S}) we arrive at the result for the flux dependent part
of the current
\begin{eqnarray}
I_{3}^{(\Phi)} &\rightarrow & \sqrt{1-\epsilon} I_{3}^{(\Phi)},
\end{eqnarray}
while in contrast to the long relaxation time result in
Eq. (\ref{currlongdep}), the current part $I_3^{(0)}$ is unaffected by
the dephasing. 

Importantly, the different dephasing behaviors in the two regimes of
probe relaxation time are clearly manifested in the pumped current. In
the long time regime the suppression of the current in the left arm
$I_{3L}^{(0)}$ leads to that the measured current only depends on the
time for ballistic propagation in the right arm, see
Eq. (\ref{I30}). As a consequence, the beating pattern in the
frequency dependence of the pumped current (see Fig. \ref{fig3}) is
suppressed on increasing dephasing. In the short time regime there is
no such suppression.

\section{Conclusions} 
 
We have investigated the pumped currents in a MZI implemented in a
conductor in the quantum Hall regime. The motivation for our
investigation was twofold. First, a MZI is the most elementary
interferometer, due to the absence of closed electronic orbits. In our
proposal the pumped current in the MZI is moreover operated solely by
modulating the potential at the two QPCs. This makes pumping in the
MZI both fundamentally important and experimentally
achievable. Second, recent experiments \cite{Ji03,Neder06,Litvin} on
transport in a voltage biased MZI has demonstrated the relevance of
dephasing and raised a number of questions on the coherence properties
of MZIs. Working in the adiabatic pumping regime makes it possible to
investigate these coherence properties close to equilibrium, keeping
dephasing at a minimum.
 
The dependence of the current on pumping frequency, pumping strength,
temperature and lengths of the arms of the MZI were investigated. The
two parts of the current, the flux dependent and the flux independent
ones, were demonstrated to depend in a qualitatively different way on
frequency and temperature. The two current parts also showed a
different sensitivity to dephasing, introduced by coupling a voltage
probe to one of the interferometer arms. The flux dependent current
was successively suppressed for increasing coupling, while only the
part of the flux independent current flowing in the arm connected to
the probe was sensitive to dephasing in the limit of long charge
relaxation time of the probe. We also demonstrated that rectification
effects, preventing an unambiguous demonstration of quantum pumping of
current, are absent in the MZI when working with semitransparent beam
splitters in the adiabatic, weak pumping regime.
   
In a broader perspective, a better understanding and control of
coherence properties of edge state transport is important for a
successful realization of two-particle Hanbury Brown Twiss
interferometers,\cite{Sam04} entanglement production
\cite{Been04,Sam04} and quantum state transfer \cite{Stace05} in
quantum Hall systems. In the context of entanglement, an unambiguous
demonstration of quantum pumping in the MZI also opens up for schemes
for entanglement generation based on quantum pump
effects.\cite{SamButt05,BeenTitov05,Mizel,Lesovik,MB06}
  
\section{Acknowledgements}
The authors would like to thank M. B\"uttiker for suggesting the
problem and providing important comments on the work. We also thank
M. Polianski for a critical reading of the manuscript. S.-W.V.C. also
thanks C.S. Tang and C.S. Chu for comments. S.-W.V.C. is supported by
the National Science Council of Taiwan under the Grant Nos.
NSC94-2112-M-009-017, MOE ATU Program, and NSC 93-2119-M-007-002
(NCTS). M.M. appreciates the support of the National Center for
Theoretical Sciences, Hsinchu, Taiwan where part of this work was
done. P.S. acknowledges support of the swedish VR.


\begin{thebibliography}{99}
\bibitem{Been91} C.W.J. Beenakker and H. van Houten, Solid State Physics
\textbf{44}, 1 (1991).

\bibitem{Ji03} Y. Ji, Y. Chung, D. Sprinzak, M. Heilblum, D. Mahalu, and H.
Shtrikman, Nature \textbf{422}, 415 (2003).

\bibitem{Neder06} I. Neder, M. Heiblum, Y. Levinson, D. Mahalu, and V.
Umansky, Phys. Rev. Lett. \textbf{96}, 016804 (2006).

\bibitem{Litvin}
L. V. Litvin, H.-P. Tranitz, W. Wegscheider, and C. Strunk, cond-mat/0607758.

\bibitem{Seelig01} G. Seelig and M. B\"{u}ttiker, Phys. Rev. B \textbf{64},
245313 (2001).

\bibitem{Marquardt04a} F. Marquardt and C. Bruder,
Phys. Rev. Lett. \textbf{92}, 56805 (2004).

\bibitem{Marquardt04b} F. Marquardt and C. Bruder, Phys. Rev. B
\textbf{70}, 125305 (2004).

\bibitem{Forster05} H. F\"{o}rster, S. Pilgram, and M. B\"{u}ttiker,
Phys.  Rev. B \textbf{72}, 075301 (2005).

\bibitem{Marquardt05} F. Marquardt, Europhys. Lett. \textbf{72}, 788
  (2005).

\bibitem{Vane05} V. S.-W. Chung, P. Samuelsson, and M. B\"{u}ttiker, Phys.
Rev. B \textbf{72}, 125320 (2005).

\bibitem{Pilgram05} S. Pilgram, P. Samuelsson, H. F\"{o}rster, and
M. B\"{u}ttiker, Phys. Rev. Lett. {\bf 97}, 066801 (2006).

\bibitem{Forster06} H. F\"{o}rster, P. Samuelsson, S. Pilgram, and
M. B\"{u}ttiker, cond-mat/0609544.

\bibitem{Sukh06}
E.V. Sukhorukov and V.V. Cheianov, cond-mat/0609288.

\bibitem{Thouless}
D. J. Thouless, Phys. Rev. B {\bf 27}, 6083 (1983).

\bibitem{BPT} M. B\"uttiker, H. Thomas, and A. Pretre, Z. Phys. B: Condens. Matter {\bf 94}, 133 (1994).

\bibitem{Spivak} B. Spivak, F. Zhou, and M. T. Beal Monod,  Phys. Rev. B 51, 13226 (1995).

\bibitem{Brouwer98} P. W. Brouwer, Phys. Rev. B \textbf{58}, R10135 (1998).

\bibitem{Zhou} F. Zhou, B. Spivak, and B. Altshuler, Phys.
Rev. Lett. \textbf{82}, 608 (1999).

\bibitem{Shutenko} T. A. Shutenko, I. L. Aleiner, and B. L. Altshuler
Phys. Rev. B {\bf 61}, 10366 (2000).

\bibitem{AEGS00} J. E. Avron, A. Elgart, G. M. Graf, and L. Sadun, Phys.
Rev. B \textbf{62}, R10618 (2000).

\bibitem{VAA01} M. G. Vavilov, V. Ambegaokar, and I. L. Aleiner, Phys. Rev.
B \textbf{63}, 195313 (2001)

\bibitem{WWG02} B. Wang, J. Wang, and H. Guo, Phys. Rev. B \textbf{65}, 073306
(2002).

\bibitem{MB02} M. Moskalets and M. B\"{u}ttiker, Phys. Rev. B. \textbf{66},
035306 (2002).

\bibitem{PB03} M. L. Polianski and P. W. Brouwer, J. Phys. A: Math. Gen. 
\textbf{36}, 3215 (2003).

\bibitem{AEGS04} J. E. Avron, A. Elgart, G. M. Graf, and L. Sadun, J. Stat.
Phys. \textbf{116}, 425 (2004).

\bibitem{MMLM04} M. Martinez-Mares, C. H. Lewenkopf, and E. R. Mucciolo,
Phys. Rev. B. \textbf{69}, 085301 (2004).

\bibitem{ZLCMcK04} H.-Q. Zhou, U. Lundin, S. Y. Cho, and R. H. McKenzie,
Phys. Rev. B \textbf{69}, 113308 (2004).

\bibitem{SH04} D. Shin and J. Hong, Phys. Rev. B. \textbf{70}, 073301
(2004).

\bibitem{CTCC04} S.W. Chung, C.S. Tang, C.S. Chu, and C.Y. Chang,
Phys. Rev.  B \textbf{70}, 085315 (2004).

\bibitem{GTFH05} M. Governale, F. Taddei, R. Fazio and F.W.J. Hekking,
Phys.  Rev. Lett. \textbf{95}, 256801 (2005).

\bibitem{CKS05} D. Cohen, T. Kottos, and H. Schanz, Phys. Rev. E
\textbf{71} , 035202(R) (2005).

\bibitem{Benjamin06}
C. Benjamin, Eur. Phys. J. B, {\bf 52}, 403 (2006). 

\bibitem{SMCG99} M. Switkes, C. M. Marcus, K. Campman, and
A. C. Gossard, Science \textbf{283}, 1905 (1999).

\bibitem{DiCarlo03}
L. DiCarlo, C.M. Marcus, and J.S. Harris, Jr, Phys. Rev. Lett. \textbf{91}, 246804 (2003).

\bibitem{Citro06}
R. Citro and F. Romeo, Phys. Rev. B {\bf 73}, 233304 (2006).

\bibitem{Kim06} S.K. Kim, K.K. Das, and A. Mizel, cond-mat/0609601.

\bibitem{SamButt05} P. Samuelsson and M. B\"{u}ttiker, Phys. Rev. B
\textbf{71}, 245317 (2005).

\bibitem{MosButt02} M. Moskalets and M. B\"{u}ttiker, Phys. Rev. B
\textbf{66}, 205320 (2002).

\bibitem{MosButt04} M. Moskalets and M. B\"{u}ttiker, Phys. Rev. B
\textbf{69}, 205316 (2004).

\bibitem{Kim}
S.W. Kim,  Phys. Rev. B \textbf{66}, 235304 (2002).

\bibitem{Brouwer01} P. W. Brouwer, Phys. Rev. B \textbf{63}, 121303(R)
(2001).

\bibitem{Butt92} M. B\"{u}ttiker, Phys. Rev. B, \textbf{46}, 12485 (1992).

\bibitem{4} M. Moskalets and M. B\"{u}ttiker, Phys. Rev. B \textbf{72},
035324 (2005).

\bibitem{Fertig88} H. A. Fertig, Phys. Rev. B \textbf{38}, 996 (1988).

\bibitem{BM06} M. B\"{u}ttiker and M. Moskalets, Lecture Notes in Physics 
\textbf{690}, 33 (2006).

\bibitem{Christen} T. Christen and M. B\"uttiker, Europhys. Lett. {\bf 35}, 523 (1996).

\bibitem{PB98} M. H. Pedersen and M. B\"uttiker, Phys. Rev. B \textbf{58},
12993 (1998).

\bibitem{Butt86} M. B\"{u}ttiker, Phys. Rev. B \textbf{33}, 3020 (1986).

\bibitem{Butt88} M. B\"{u}ttiker, IBM J. Res. Dev. \textbf{32}, 63 (1988).

\bibitem{BlanterButt00} Ya. M. Blanter and M. B\"{u}ttiker, Phys. Rep.
{\bf 336}, 1 (2000).

\bibitem{TexierButt92} C. Texier and M. B\"{u}ttiker, Phys. Rev. B
\textbf{62}, 7454 (2000).

\bibitem{Ober06} S. Oberholzer, E. Bieri, C. Sch\"{o}nenberger, M. Giovannini,
and J. Faist, Phys. Rev. Lett. \textbf{96}, 46804 (2006).

\bibitem{MosButt01} M. Moskalets and M. B\"{u}ttiker, Phys. Rev. B
\textbf{64}, 201305(R) (2001).

\bibitem{Cremers02} J.N.H.J. Cremers and P.W. Brouwer, Phys. Rev. B {\bf 65}, 115333 (2002).

\bibitem{PoliButt05} M. L. Polianski, P. Samuelsson, and M. B\"{u}ttiker,
Phys. Rev. B \textbf{72}, 161302(R) (2005).

\bibitem{PoliBrouw01} M.L. Polianski, and P.W. Brouwer, Phys. Rev. B \textbf{64} 075304 (2001). 

\bibitem{Sam04} P. Samuelsson, E.V. Sukhorukov and M. B\"uttiker, Phys. Rev. Lett. {\bf 92}, 026805 (2004).

\bibitem{Been04} C.W.J. Beenakker, C. Emary, M. Kindermann, and J.L. van Velsen, Phys. Rev. Lett. {\bf 91}, 147901 (2003).

\bibitem{Stace05} T. M. Stace, C. H. W. Barnes, and G. J. Milburn, Phys. Rev. Lett. {\bf 93}, 126804 (2004).

\bibitem{BeenTitov05} C.W.J. Beenakker, M. Titov, and B. Trauzettel, Phys.
Rev. Lett. \textbf{94}, 186804 (2005).

\bibitem{Mizel}
K. K. Das, S. Kim, and A. Mizel, Phys. Rev. Lett. {\bf 97}, 096602 (2006).

\bibitem{Lesovik}
A. V. Lebedev, G. B. Lesovik, and G. Blatter, Phys. Rev. B {\bf 72}, 245314 (2005).

\bibitem{MB06} M. Moskalets and M. Buttiker, Phys. Rev. B {\bf 73}, 125315
(2006).

\end{thebibliography}
\end{document}